\documentclass[twocolumn,showpacs,superscriptaddress,preprintnumbers,amsmath,amssymb,prc]{revtex4-1}

\usepackage{graphicx}
\usepackage{dcolumn}
\usepackage{bm}
\usepackage{indentfirst} 
\usepackage{color}
\usepackage{CJK}
\usepackage{booktabs}
\usepackage{float}
\usepackage{tikz}
\usepackage[colorlinks,linkcolor=blue,urlcolor=blue,citecolor=blue]{hyperref}

\makeatletter
\renewcommand{\fnum@figure}{\textbf{Fig.} \thefigure.\@gobble}
\makeatother
\newcommand{\reffig}[1]{Fig. \ref{#1}}
\newcommand{\reftab}[1]{Table \ref{#1}}
\newcommand{\refequ}[1]{Eq. (\ref{#1})}

\begin{document}

\title{Fudan Multi-purpose Active TArget Time Projection Chamber (fMeta-TPC) for Photonnuclear Reaction Experiments}\thanks{This work is supported in part by National Key R\&D Program of China Nos. 2022YFA1602402, 2020YFE0202001, 2023YFA1606900 and the National Natural Science Foundation of China (NSFC) under Grant Nos. 12235003, 11835002, 11705031, 12275053, and 12147101.}

\author{Huang-Kai Wu}
\affiliation{Key Laboratory of Nuclear Physics and Ion-beam Application (MOE), Institute of Modern Physics, Fudan University, Shanghai 200433, China}
\author{Xi-Yang Wang}
\affiliation{Key Laboratory of Nuclear Physics and Ion-beam Application (MOE), Institute of Modern Physics, Fudan University, Shanghai 200433, China}
\author{Yu-Miao Wang}
\affiliation{Key Laboratory of Nuclear Physics and Ion-beam Application (MOE), Institute of Modern Physics, Fudan University, Shanghai 200433, China}
\author{You-Jing Wang}
\affiliation{Key Laboratory of Nuclear Physics and Ion-beam Application (MOE), Institute of Modern Physics, Fudan University, Shanghai 200433, China}
\author{De-Qing Fang}
\affiliation{Key Laboratory of Nuclear Physics and Ion-beam Application (MOE), Institute of Modern Physics, Fudan University, Shanghai 200433, China}
\author{Wan-Bing He}
\affiliation{Key Laboratory of Nuclear Physics and Ion-beam Application (MOE), Institute of Modern Physics, Fudan University, Shanghai 200433, China}
\author{Wei-Hu Ma}
\affiliation{Key Laboratory of Nuclear Physics and Ion-beam Application (MOE), Institute of Modern Physics, Fudan University, Shanghai 200433, China}
\author{Xi-Guang Cao}
\affiliation{Shanghai Institute of Applied Physics, Chinese Academy of Sciences, Shanghai 201800, China}
\affiliation{University of Chinese Academy of Sciences, Beijing 100049, China}
\affiliation{Shanghai Advanced Research Institute, Chinese Academy of Sciences, Shanghai 201210, China}

\author{Chang-Bo Fu}
\email[Corresponding author, ]{Chang-Bo Fu, Key Laboratory of Nuclear Physics and Ion-beam Application (MOE), Institute of Modern Physics, Fudan University, Shanghai 200433, China, cbfu@fudan.edu.cn.}
\affiliation{Key Laboratory of Nuclear Physics and Ion-beam Application (MOE), Institute of Modern Physics, Fudan University, Shanghai 200433, China}
\author{Xian-Gai Deng}
\email[Corresponding author, ]{Xian-Gai Deng, Key Laboratory of Nuclear Physics and Ion-beam Application (MOE), Institute of Modern Physics, Fudan University, Shanghai 200433, China, xiangai_deng@fudan.edu.cn.}
\affiliation{Key Laboratory of Nuclear Physics and Ion-beam Application (MOE), Institute of Modern Physics, Fudan University, Shanghai 200433, China}
\affiliation{Shanghai Research Center for Theoretical Nuclear Physics, NSFC and Fudan University, Shanghai 200438, China}
\author{Yu-Gang Ma}
\email[Corresponding author, ]{Yu-Gang Ma, Key Laboratory of Nuclear Physics and Ion-beam Application (MOE), Institute of Modern Physics, Fudan University, Shanghai 200433, China, mayugang@fudan.edu.cn.}
\affiliation{Key Laboratory of Nuclear Physics and Ion-beam Application (MOE), Institute of Modern Physics, Fudan University, Shanghai 200433, China}
\affiliation{Shanghai Research Center for Theoretical Nuclear Physics, NSFC and Fudan University, Shanghai 200438, China}

\begin{abstract}
Active Target Time Projection Chambers (AT-TPCs) are state-of-the-art tools in the field of low-energy nuclear physics, particularly suitable for experiments using low-intensity radioactive ion beams or gamma rays. The focus is on the study of the photonuclear reaction with the Laser Compton Scattering (LCS) gamma source, especially for the decay of the highly excited $\alpha$-cluster state. The Fudan Multi-purpose Active Target Time Projection Chamber (fMeta-TPC) with 2048 channels has been developed to study $\alpha$-clustering nuclei. In this work, the design of fMeta-TPC is described and a comprehensive evaluation of its offline performance is performed by ultraviolet (UV) laser and $^{241}$Am $\alpha$ source. 
The result shows that the intrinsic angular resolution of the detector is within 0.30$^{\circ}$ and has an energy resolution of 6.85\% for 3.0 MeV $\alpha$ particles. The gain uniformity of the detector is about 10\% (RMS/Mean), tested by the $^{55}$Fe X-ray source. 
\end{abstract}


\maketitle

\section{Introduction}
Recently, the establishment of the Shanghai Laser Electron Gamma Source (SLEGS) beamline at the Shanghai Synchrotron Radiation Facility (SSRF) has provided a promising platform for studying MeV-level photonuclear reactions in China \cite{bib:1}. The LCS gamma source offers advantages over traditional $\gamma$-ray sources, including quasi-monoenergetic, high brightness, and high polarization. Photonuclear reactions, characterized by simple reaction mechanisms and clean final-state products, serve as effective probes of nuclear structure and measurements of key reaction rates in nuclear astrophysics \cite{bib:2,bib:3,bib:4,bib:5,bib:6,bib:7,bib:8}.
However, most important photonuclear reaction experiments suffer from the drawback of small reaction cross sections and low energy reaction products.

The AT-TPCs play a crucial role in low-energy nuclear physics and are considered as novel and powerful detector tools \cite{bib:9,bib:10}. 
They have been widely used in various fields, such as the ACTAR TPC of the Grand Accélérateur National d'Ions Lourds (GANIL) for the study of shell evolution \cite{bib:11}, the O-TPC of the University of Warsaw for the study of cluster structure \cite{bib:12}, nuclear astrophysics \cite{bib:13} and exotic decay \cite{bib:14}, MAIKo of Kyoto University for the study of shell evolution and cluster structure \cite{bib:15,bib:16}, and TexAT of Texas A\&M University for the study of shell evolution and nuclear astrophysics \cite{bib:17}. 
AT-TPCs are designed to use different gases simultaneously as targets and detectors, providing nearly 4$\pi$ solid angle coverage and a low energy detection threshold. Furthermore, since the target itself serves as the detector, a thicker target does not compromise energy resolution and detection efficiency, which is particularly beneficial for photonuclear reactions and low-intensity beam experiments.

In addition, AT-TPCs are particularly useful instruments for nuclear cluster studies.
The study of cluster structure in light nuclei is a prominent research frontier in nuclear physics \cite{bib:18,bib:19,bib:20,bib:21,bib:22,Yang,Fang,Ye}. A well-known example is the Hoyle state of $^{12}$C, first posited by F. Hoyle in 1953 to explain nucleosynthesis in stars \cite{bib:23}. Although research using various methods has indicated the presence of significant cluster components in the ground and low-lying excited states of light $\alpha$-conjugate nuclei \cite{bib:12,bib:24,bib:25}, the exact properties and configurations of $\alpha$ clusters remain elusive. Questions remain regarding the distinction between $\alpha$ cluster states and free $\alpha$ particles, and whether $\alpha$ clusters adopt a Bose-Einstein Condensate (BEC) state \cite{bib:26,bib:27,bib:28} or specific geometric configurations \cite{bib:29,bib:30,bib:31}. Thus, the properties and structures of clusters in nuclei such as $\alpha$-conjugated nuclei like $^{12}$C, $^{16}$O, and $^{20}$Ne, or non$\alpha$-conjugated nuclei such as $^{6}$Li and $^9$Be, including the Hoyle state in $^{12}$C and analogous Hoyle-like cluster states in other $\alpha$-conjugated nuclei \cite{bib:28,bib:32,bib:33}, is an important open question.

Considering these advantages, an AT-TPC, namely the Fudan Multi-purpose Active Target Time Projection Chamber (fMeta-TPC), has been built. 
The fMeta-TPC was designed and constructed with a special focus on the study of photonuclear reactions, especially the properties of $\alpha$-clusters in the excited states of light nuclei.

The article is organized as follows: In section \ref{sec:fMeta-TPC} we describe the design of the fMeta-TPC. 
The features of the readout board are presented in section \ref{sec:Micromegas}, 
and the design and simulation of the electronic field uniformity of the field cage is presented in section \ref{sec:Fieldcage}. 
An overview of the electronic system and its basic performance characteristics is given in section \ref{sec:Electronic}. 
Offline performance tests of the detector, covering electron drift velocity, electronic field homogeneity, angular resolution and energy resolution of the TPC, are detailed in section \ref{sec:performance}. Finally, a summary is given in section \ref{Summary}.

\section{Design of fMeta-TPC}
\label{sec:fMeta-TPC}
The fMeta-TPC was developed at Fudan University, Shanghai, China. The four main components of the fMeta-TPC are the gas chamber, the anode pad plane, the field cage and the electronic system. A schematic view is shown in \ref{fig:gas_chamber}.

\subsection{Gas chamber}
\begin{figure}[!htb]
    \centering
    \includegraphics[width=0.9\linewidth]{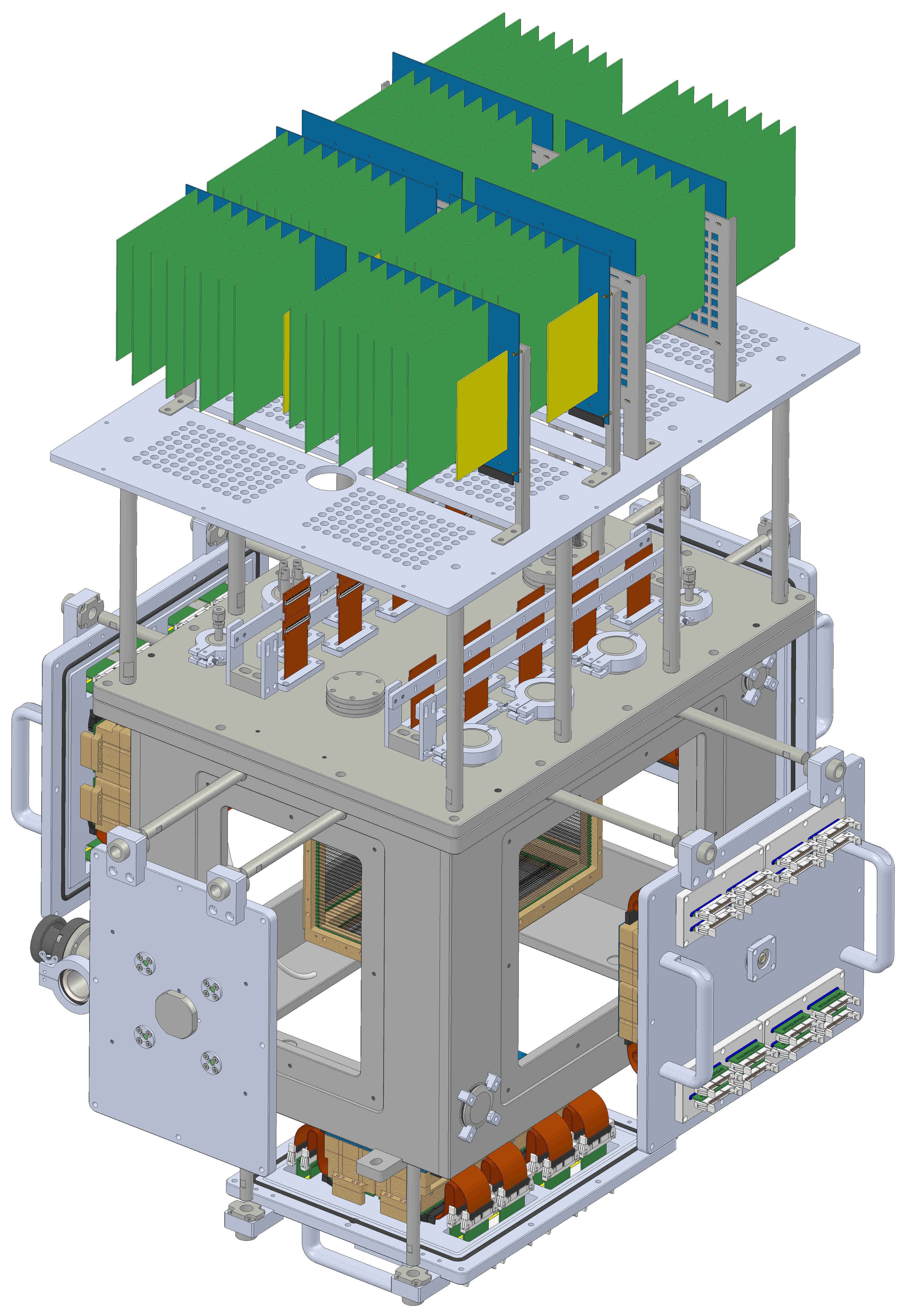}
    \caption{(Color Online) 3D Schematic view of fMeta-TPCs system. }
    \label{fig:gas_chamber}
\end{figure}

As shown in  \reffig{fig:gas_chamber}, fMeta-TPC is housed in a cubic stainless steel chamber with a volume of 600 (L) $\times$ 450 (W) $\times$ 475 (H) mm$^3$. This chamber is equipped with five removable flanges, located at the bottom and along the four sides, to allow easy mounting and testing of detectors. On the beam injection side, the flange has a 15 mm diameter circular entrance window and four quartz windows. A 0.3 mm thick Kapton foil, capable of withstanding a pressure differential of 1 atm, is used to maintain isolation between the gas volume and the external environment. Due to the low stopping power of the gas, the charged particle with several MeV energy can easily escape from the filed cage. In view of future experimental plans on rare isotope beams, the remaining flanges are designed to support the auxiliary detectors. For example, the $\Delta$E-E telescopes are composed of a double sided silicon strip detector (DSSSD) and a 50 mm thick CsI(Tl) detector, which has been designed but not introduced in this work.

\subsection{Resistive Micromegas}
\label{sec:Micromegas}
\begin{figure}[!htb]
    \centering
    \includegraphics[width=\linewidth]{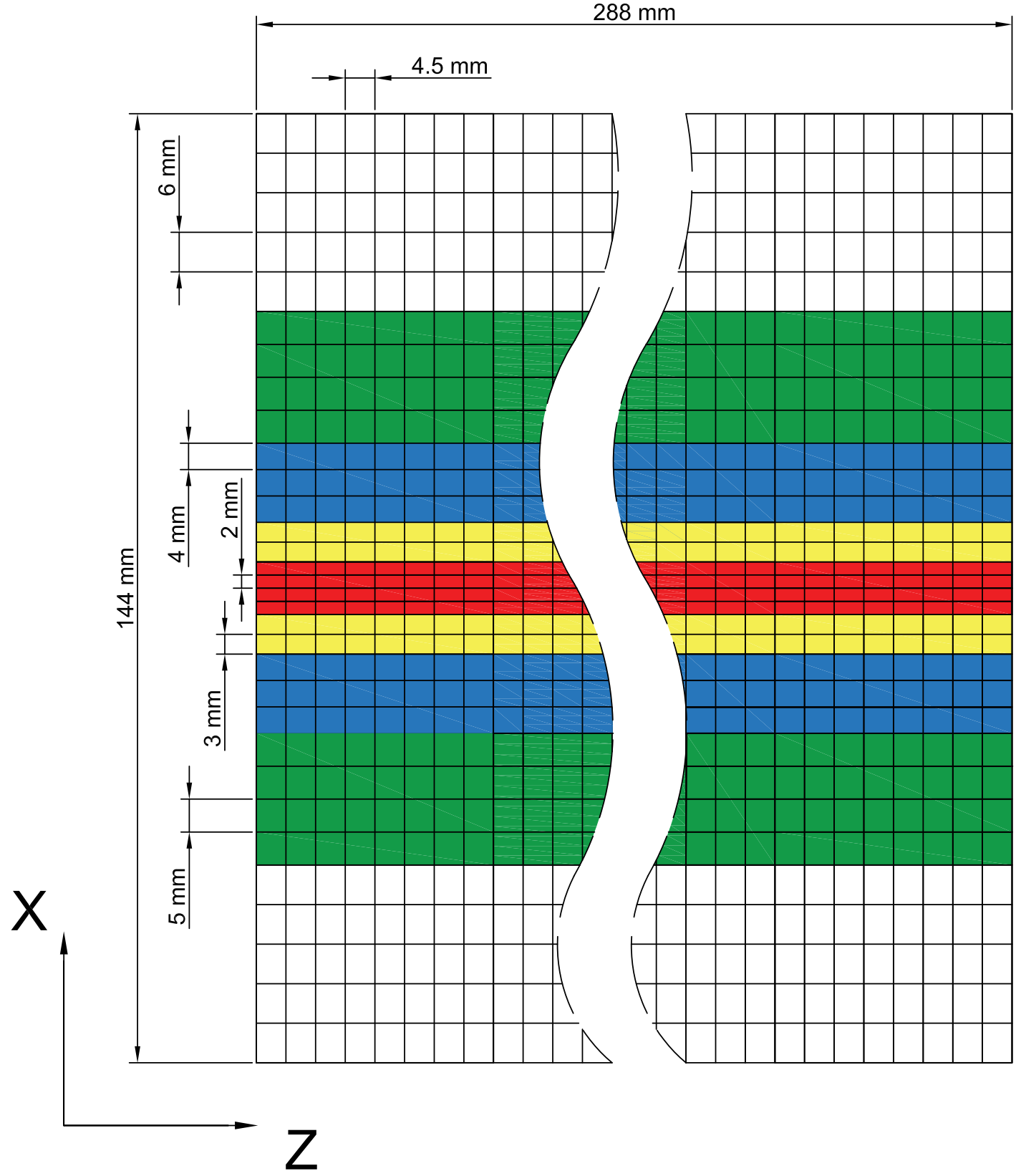} 
    \quad
    \includegraphics[width=\linewidth]{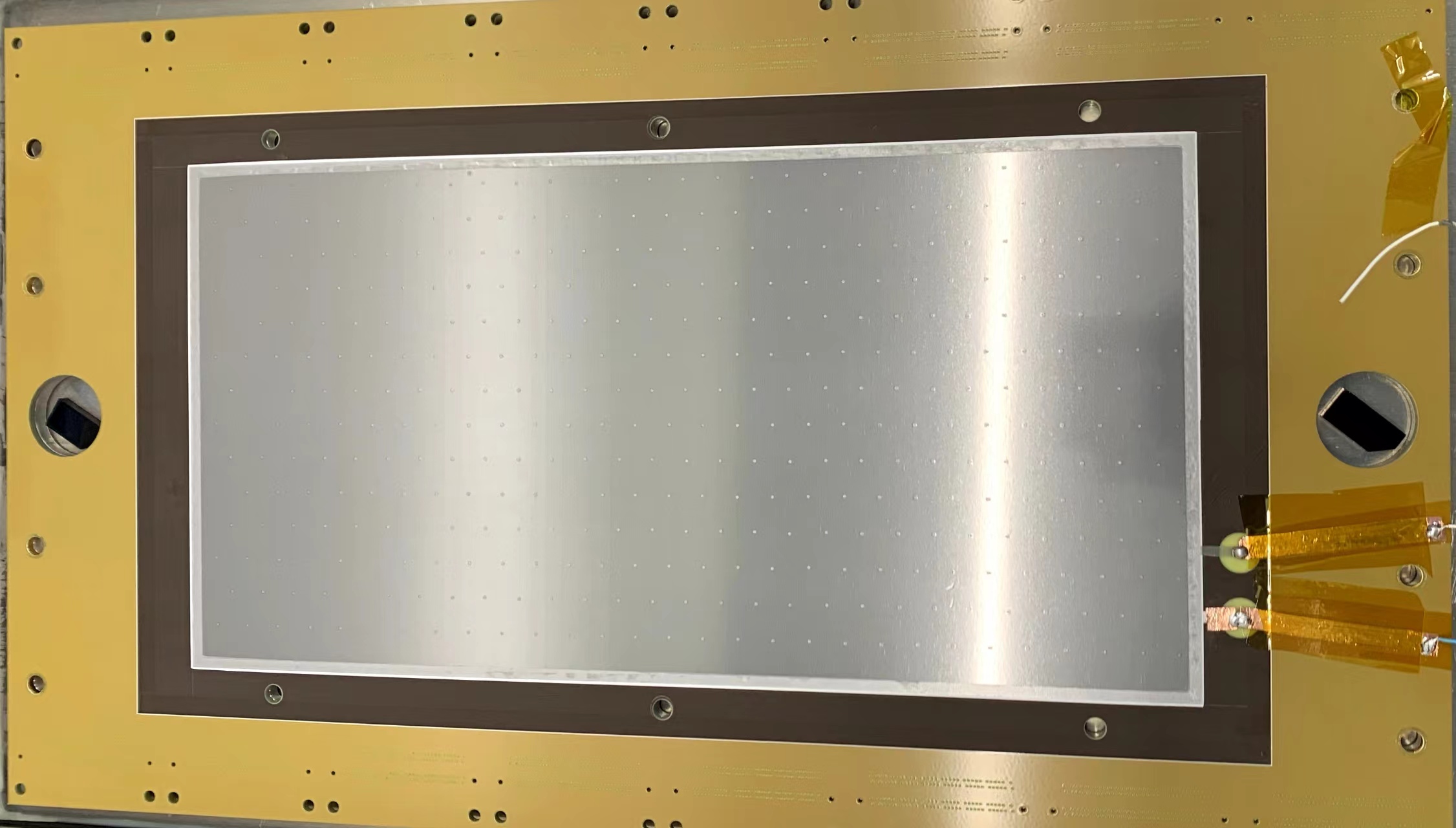}
    \caption{(Color Online) 
    Top panel: The schematic view of the sensitive area ($144\times288 $ mm$^2$) of the micromagas plane. The sensitive area is divided into $32\times64$ rectangular pixels of different sizes. The pixel sizes in the $x$-direction are 2 mm, 3 mm, 4 mm, 5 mm, and 6 mm, while in the $z$-direction it is kept constant at 4.5 mm. Bottom panel: A photo of the Micromegas detector.
    }
    \label{fig:micromegas}
\end{figure}

As shown in the \reffig{fig:micromegas}, the anode pad plane coupled with the field cage is housed by the top flange. And its voltages are supplied through safe high voltage (SHV) connectors located on the top flange. In this work, the Micromegas resistive detector was used for electron amplification and collection. The photograph and the schematic view of the pad plane are shown in \reffig{fig:micromegas}. 
The Micromegas readout pad was fabricated at the University of Science and Technology of China using the thermal bonding technique \cite{bib:34}.
The mesh of the Micromegas has 325 lines per inch (LPI) with 23 $\mu$m diameter wires and a 49\% opening rate. The avalanche gap between the mesh and the surface of the printed circuit board (PCB) is about 100 $\mu$m. And a 5 mm thick aluminum plate was glued and screwed to the back of the PCB to increase its mechanical rigidity. 

The anode pad plane has a sensitive area of $144\times288$ mm$^2$, consisting of 2048 readout channels. Due to the characteristics of photodisintegration reactions, the sensitive area is segmented into $32\times64$ rectangle pixels, which increase in size from the inner to the outer region, as shown in \reffig{fig:micromegas}. The pixel size in the $x$-direction (short side) increases from 2 mm to 6 mm, while it remains constant at 4.5 mm in the $z$-direction (long side). 
The smaller pixels in the inner region are designed to improve the angular resolution of the detector for short range particles. And the larger pixels in the outer region make the anode pad plane to cover a larger sensitive area with a limited number of electronic channels.
All pixels are read out through high-density (0.5 mm pitch) Hirose FX10A-140P/14-SV1(71) connectors, which contain two rows of 140 pins each. 
More details on the angular resolution tests are given in section \ref{sec:angular}.

\subsection{Field cage}
\label{sec:Fieldcage}
\begin{figure}[!htb]
    \centering
    \includegraphics[width=\linewidth]{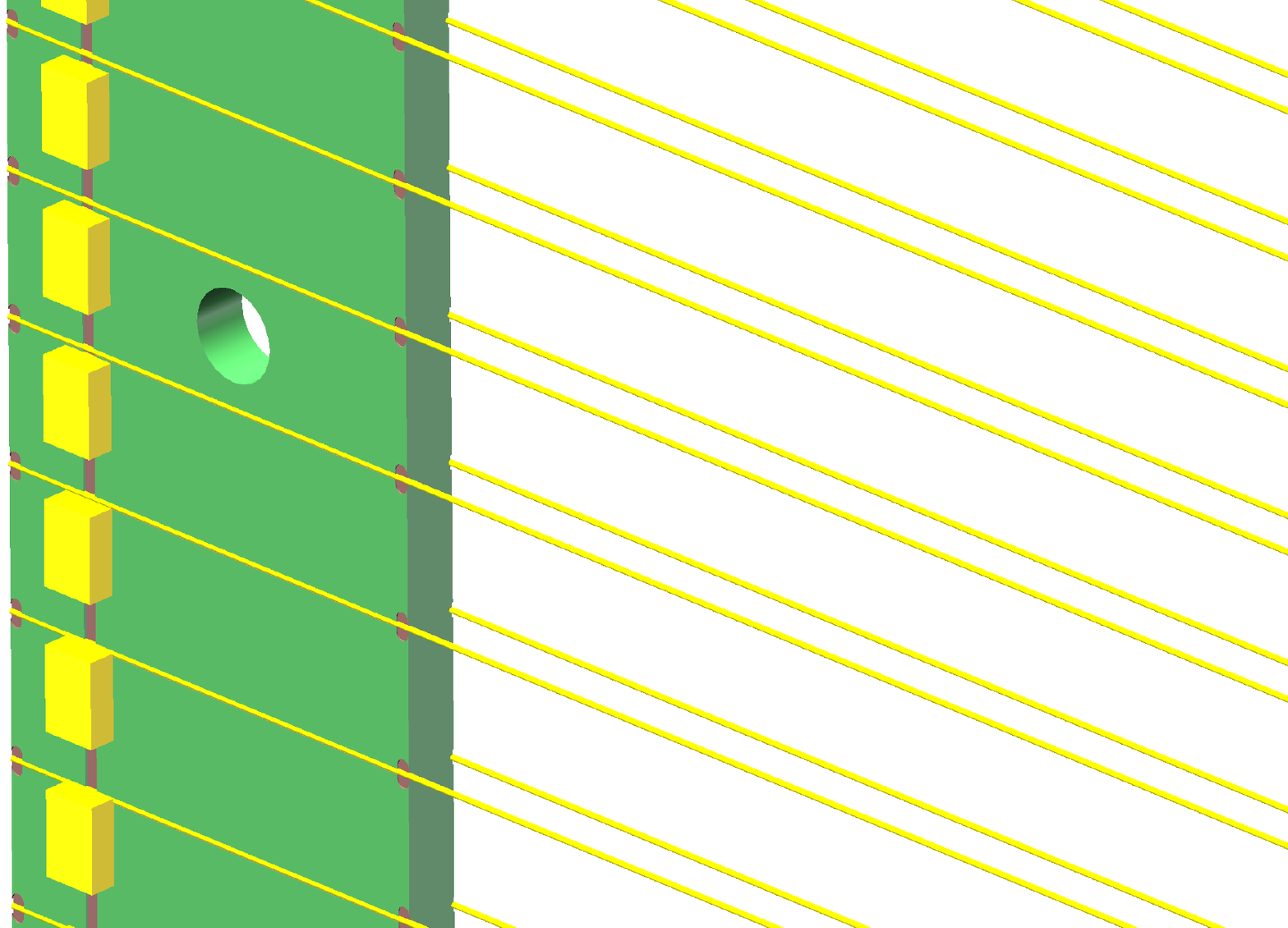}
    \caption{(Color Online) 3D drawing of the double layer gold plated tungsten wire side plane of the field cage. The yellow blocks represent 10 M$\Omega$ surface mount resistors.
    }
    \label{fig:pcb}
\end{figure}

The vertical uniform electric field within the sensitive volume of the TPC is formed by the field cage with a volume of 330 (L) $\times$ 180 (W) $\times$ 180 (H) mm$^3$. Two versions of the field cage have been developed for different experimental purposes: the  double wire-plane field cage and the PCB field cage. 

The double wire-plane field cage was designed specifically for the radioactive ion beam experiment, which is known to produce high-energy reaction products. To ensure accurate measurements, it was critical to design a field cage that was transparent to both the beam ions and the reaction products while maintaining a uniform electric field. Therefore, the double wire plane structure was chosen as the most effective approach. This design features five wire planes, which are surrounded by gold-plated tungsten wire and soldered together, as shown in \reffig{fig:pcb}. The four-sided wire planes are composed of 70 wires, each with a diameter of 50 $\mu$m and a vertical spacing of 5.08 mm. While the cathode plane is composed of 30 $\mu$m wires with 3.05 mm distance between each wire. The horizontal distance from the inner to the outer wire plane is 3 mm for all wire planes. This design makes the double-wire-plane field cage nearly 99\% optically transparent.

Another PCB field cage was designed specifically for the photonuclear reaction experiment in SLEGS. Except for the beam entrance plane, which remains the double wire plane, the other side planes of the PCB field cage are constructed using etched copper plate lines of 1 mm diameter. And a 3 mm thick copper-clad PCB with dimensions of $180\times330$ mm$^2$ serves as the cathode. The wires in the side plane are connected in series with 10 M$\Omega$ (0.1\% error) surface mount resistors for both field cages to homogeneously degrade the voltage from the micromesh (GND) to the cathode (-HV). By adjusting the voltage of the cathode and anode, different drift and avalanche field strengths can be set.

\begin{figure}[!htb]
  \begin{minipage}{\linewidth}
     \vspace{3pt}  
     \centerline{\includegraphics[width=0.9\textwidth]{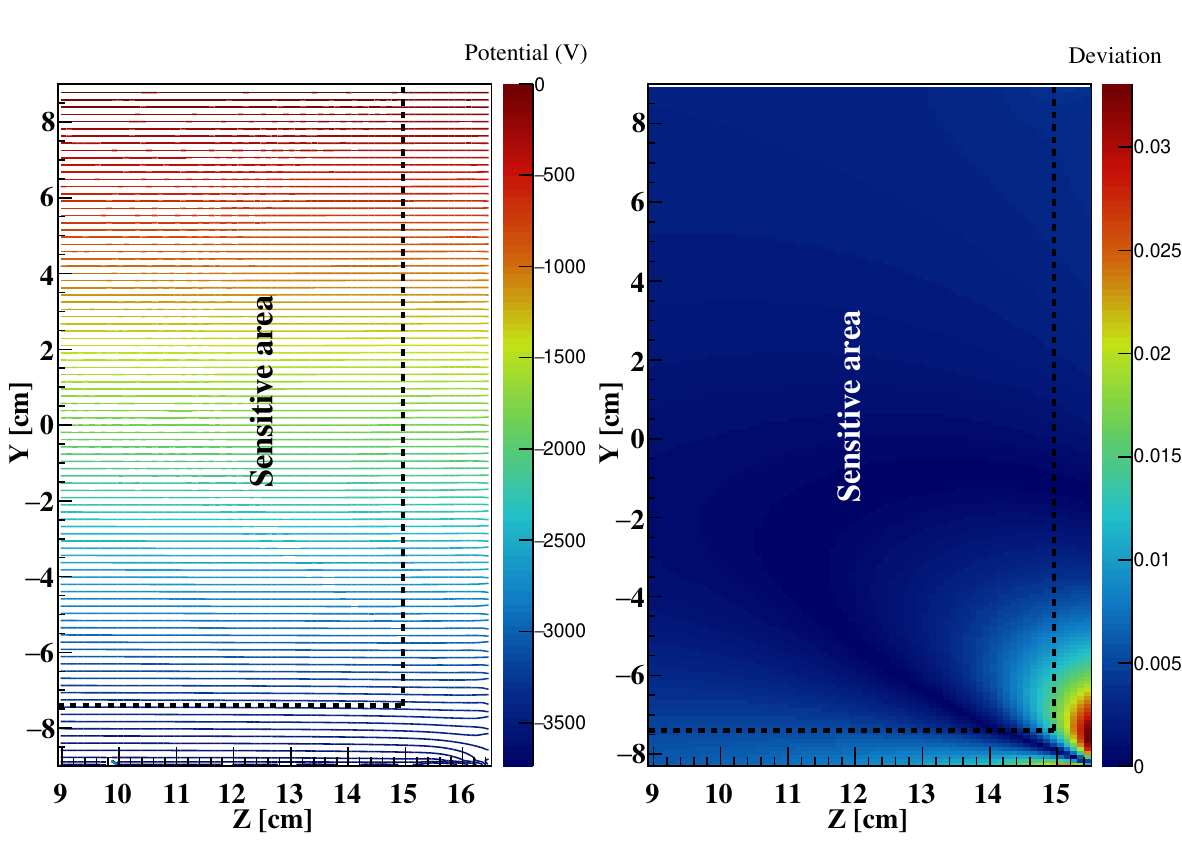}}
     \vspace{3pt}
     \centerline{\includegraphics[width=0.9\textwidth]{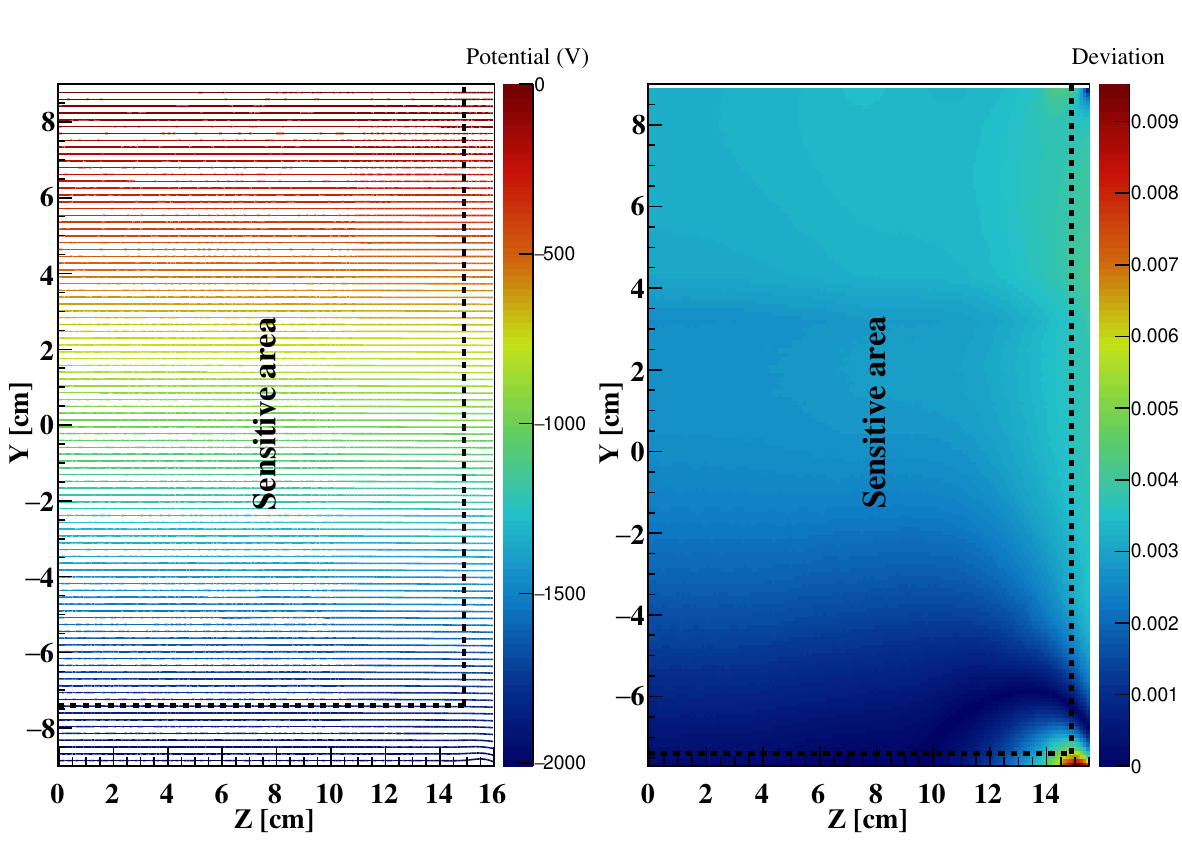}}
  \end{minipage}
 \caption{(Color Online) Simulation of the electric potential (left) and electric field (right) of the field cage. The black dashed line marks the boundary in the sensitive area of the detector. The top two panels show the simulation results of the double-wire-plane field cage, while the bottom two panels show the simulation results of the PCB field cage.}
 \label{fig:E_field}
\end{figure}

To verify the uniformity of the electric field in both field cages, a finite element calculation was performed using the neBEM code \cite{bib:35}, which is integrated into the Garfield++ code \cite{bib:36}. The simulation showed that the field cage maintains a distortion of the electric field within 3\% for the double wire-plane field cage and 1\% for the PCB field cage in the sensitive volume, as shown in \reffig{fig:E_field}.

\subsection{Readout electronic system}
\label{sec:Electronic}

\begin{table}[!htb]
    \caption{The key parameters of the readout electronic system.}
    \centering
    \label{tab:electronic}
    \begin{tabular*}{8cm} {@{\extracolsep{\fill} } ll}
        \toprule
        Parameters & Value  \\
        \midrule
        Channels               & 2048 \\
        Dead time              & 25 $\mu$s \\
        Sampling rate          & 40 MSPS \\
        Quantization accuracy  & 12 bit \\
        Sampling window width  & 25.6 $\mu$s\\
        Dynamic range          & 2 fC to 3 pC \\
        \bottomrule
    \end{tabular*}
\end{table}

In view of possible future experiments, the electronic systems of the fMeta-TPC should meet requirements such as high integration, compactness, short dead time, low power consumption and high dynamic range.
The fMeta-TPC has been designed for 2048 readout channels and future upgrades for higher spatial resolution, which places specific requirements on high integration and compactness of the readout electronics.  
High dynamic ranges of the detected energy are also required. 
For example, in reactions such as $^{16}\mathrm{O}(\gamma, \alpha) ^{12}\mathrm{C}$, the ionization energy loss ratio between $^{12}\mathrm{C}$ and $\alpha$ can be several hundred times, requiring a wide dynamic range for charge measurement in the electronics.

The key parameters of the electronic system are listed in \reftab{tab:electronic}. It has a wide dynamic range of 2 fC to 3 pC and a short dead time of 25 $\mu$s (much less than the dead time of the GET electronics) for high event rates \cite{bib:37}.

\begin{figure*}[!htb]
    \centering
    \includegraphics[width=\linewidth]{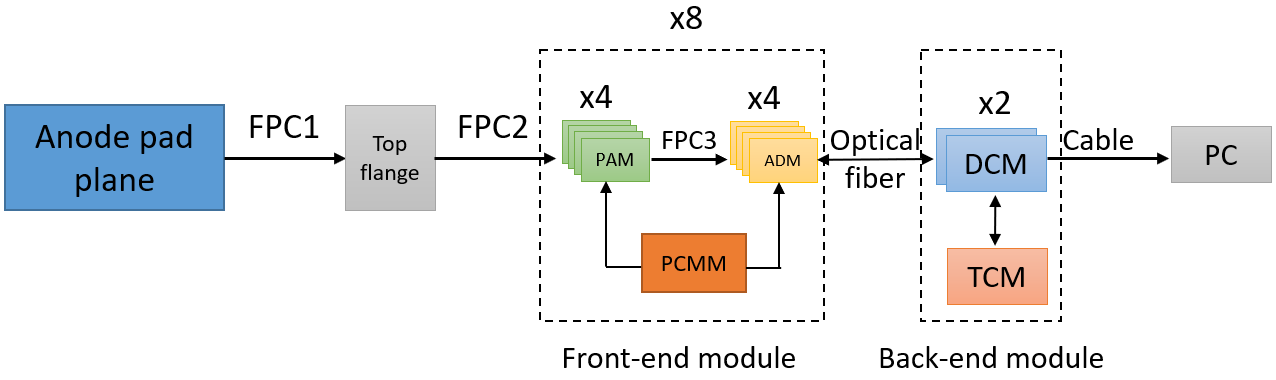}
    \caption{(Color Online) Schematic of the readout electronics system.}
    \label{fig:electronic}
\end{figure*}

As shown in in \reffig{fig:electronic}, 
the readout system is divided into two components: the front-end module and the back-end module. The front-end module, located at the top of the chamber (as shown in \reffig{fig:gas_chamber}), interfaces with the detector via two Flexible Printed Circuit (FPC) boards.

The front-end module consists of eight groups, each containing the pre-amplifier module (PAM), the analog-to-digital module (ADM), and the power clock management module (PCMM). Within each group, there are four PAM modules responsible for charge integration, four ADM modules responsible for waveform digitization, and one PCMM module responsible for supplying power to the front-end modules.

Upon receiving a signal from the detector, the PAM initiates amplification. There are 64 channels on a single PAM board, which consumes approximately 1 W of power. After amplification, the signal is transmitted to the ADM via the FPC board. The ADM then performs simultaneous digitization of 64 channels of analog signals at a sampling rate of 40 MHz and a quantization accuracy of 12 bits. Each ADM board consumes about 8 W of power. Finally, the PCMM provides power and clock distribution to the group of four PAMs and ADMs.

As for the back-end modules, it contains two parts: the Trigger Clock Module (TCM) and the Data Concentrator Module (DCM). The TCM is responsible for generating a global trigger and synchronous clock by accepting hit information and distributing the trigger and synchronous clock to the DCM. On the other hand, the DCM is responsible for data acquisition, digital filtering, buffering, and data upload to the server via FPGA. It also distributes the synchronous clock, instructions, and triggers to the front-end electronics. In our case, the back-end modules have two DCMs and one TCM. Each DCM board is connected to sixteen ADM boards via fiber optics.
\section{Performance test}
\label{sec:performance}
In this section several methods are used to characterize the detector performance. The gain variation of the Micromegas detector was tested using the $^{55}$Fe 5.9 keV X-ray source. The UV laser, which can ionize gas molecules and form a straight line in the gas, was used to study the drift velocity, field homogeneity, intrinsic angular and position resolutions of the detector. And the $^{241}$Am alpha source was used to test the energy resolution of the detector. In this work, tests were mainly performed in the Ar+$\rm CH_4$ (9:1, P10 gas), Ar+$\rm iC_4H_{10}$ (93:7) and He+$\rm CO_2$ (96:4) gas mixtures at different pressures.
\subsection{Drift velocity}

\begin{figure*}[!htb]
  \begin{minipage}{0.48\linewidth}
     \vspace{3pt}  
     \centerline{\includegraphics[width=\textwidth]{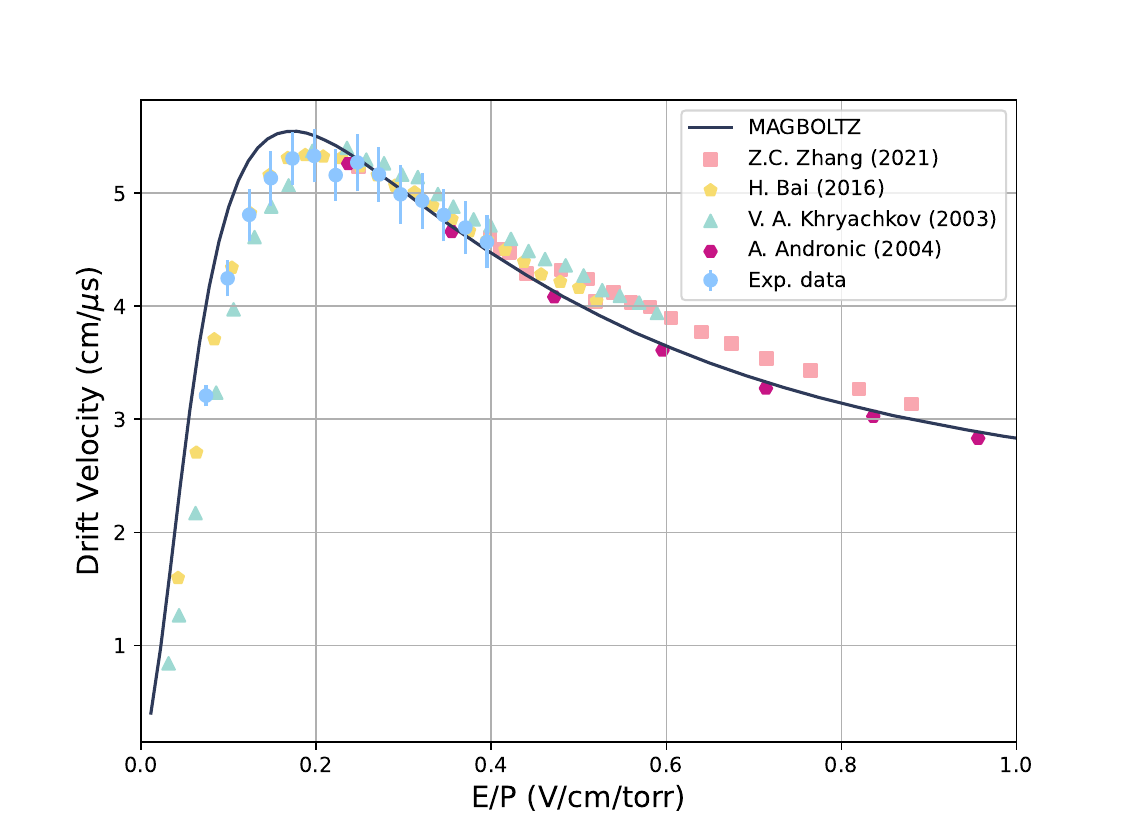}}
  \end{minipage}
  \begin{minipage}{0.48\linewidth}
     \vspace{3pt}
     \centerline{\includegraphics[width=\textwidth]{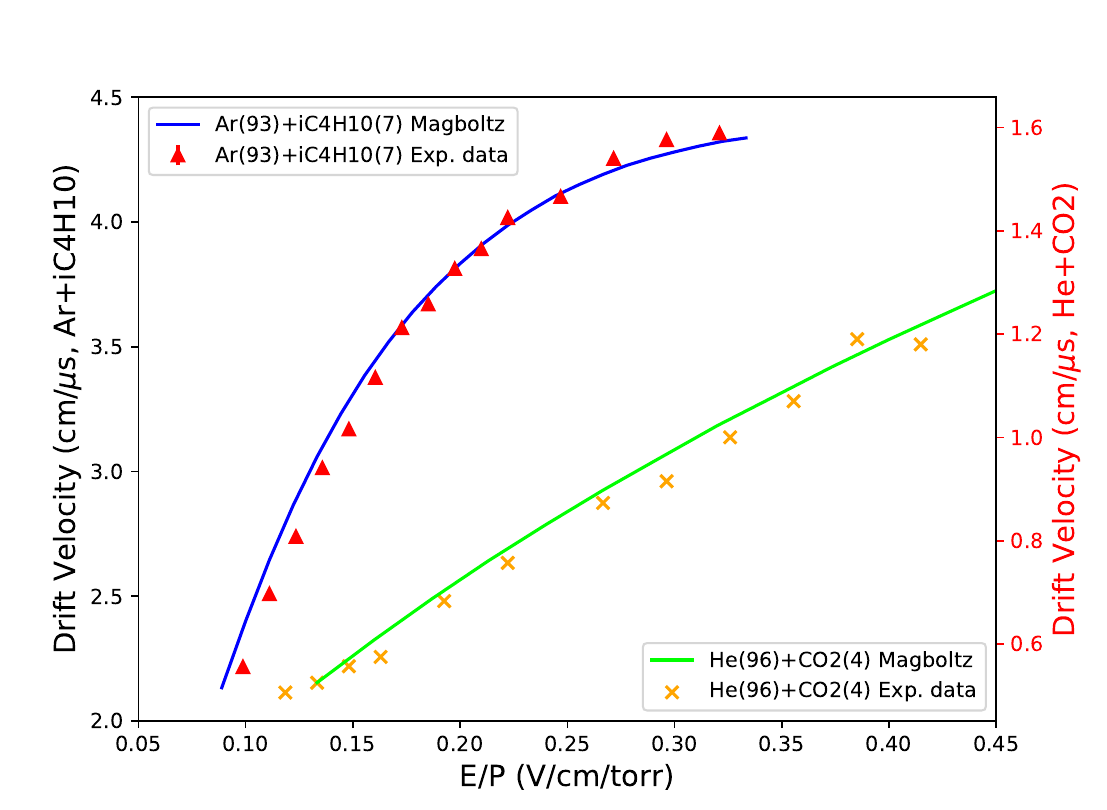}}
 \end{minipage}
 \caption{(Color Online) The electron drift velocity measured in P10 (left), Ar+$\rm iC_4H_{10}$ (93:7), and He+$\rm CO_2$ (96:4) gas mixtures (right). The measured data are compared with the calculations of MagBoltz. Reference data are taken from Z. C. Zhang \cite{bib:38}, H. Bai \cite{bib:39}, V. A. Khryachkov \cite{bib:40}, A. Andronic \cite{bib:41}.}
 \label{fig:drift_velocity}
\end{figure*}

One of the key advantages of TPC is its ability to record the 3D trajectory of charged particles. The determination of the 3D information is based on the product of the electron drift velocity and the drift time. Therefore, an accurate measurement of the electron drift velocity in the gas is crucial for particle trajectory reconstruction. In this study, the velocity was determined using a 266 nm laser with a power of 20 mW and a frequency of 7.5 kHz repetition rate. The laser entered the detector through quartz windows mounted on the front flange of the chamber, as shown in  \reffig{fig:gas_chamber}. Prior to the test, the orientation of the laser was determined based on the location of the incident point on the quartz window and the end point of the track in the field cage. 
Due to limitations imposed by the maximum processing frequency of the ADM chip and the maximum transmission capacity of the gigabit network. The data acquisition was triggered by an external 100 Hz trigger and the acquisition window is set to 25 $\mu$s. 
Considering that the pixel size remains a constant value of 4.5 mm in the $z$ direction, it becomes possible to calculate the relative differences in drift height of each pad once the laser direction is established. Therefore, one of the fired pads was chosen as a reference pad, and then the drift velocity was calculated by calculating the disparity in drift height and time with respect to this reference pad.

\reffig{fig:drift_velocity} shows the measured drift velocity in different gases compared with the theoretical calculations of MagBoltz \cite{bib:35} and other experimental results \cite{bib:38,bib:39,bib:40,bib:41}. The measured results, as shown in \reffig{fig:drift_velocity} (left), in the 600 mbar P10 gas are in good agreement with other experimental results. However, a slight deviation from the theoretical results is observed when ${\mathrm{E/P}} <$ 0.23 V/cm/torr. And the results shown in \reffig{fig:drift_velocity} (right) represent the electron drift velocity in 600 mbar Ar+$\rm iC_4H_{10}$ (93:7) and 500 mbar He+$\rm CO_2$ (96:4) mixed gases, respectively. Both measurement results agree with theoretical calculations within an uncertainty of 5\%. This indicates that our test method is applicable to the measurement of electron drift velocity.
\subsection{Drift field homogeneity}

\begin{figure*}[!htb]
  \begin{minipage}{0.48\linewidth}
     \vspace{3pt}  
     \centerline{\includegraphics[width=\textwidth]{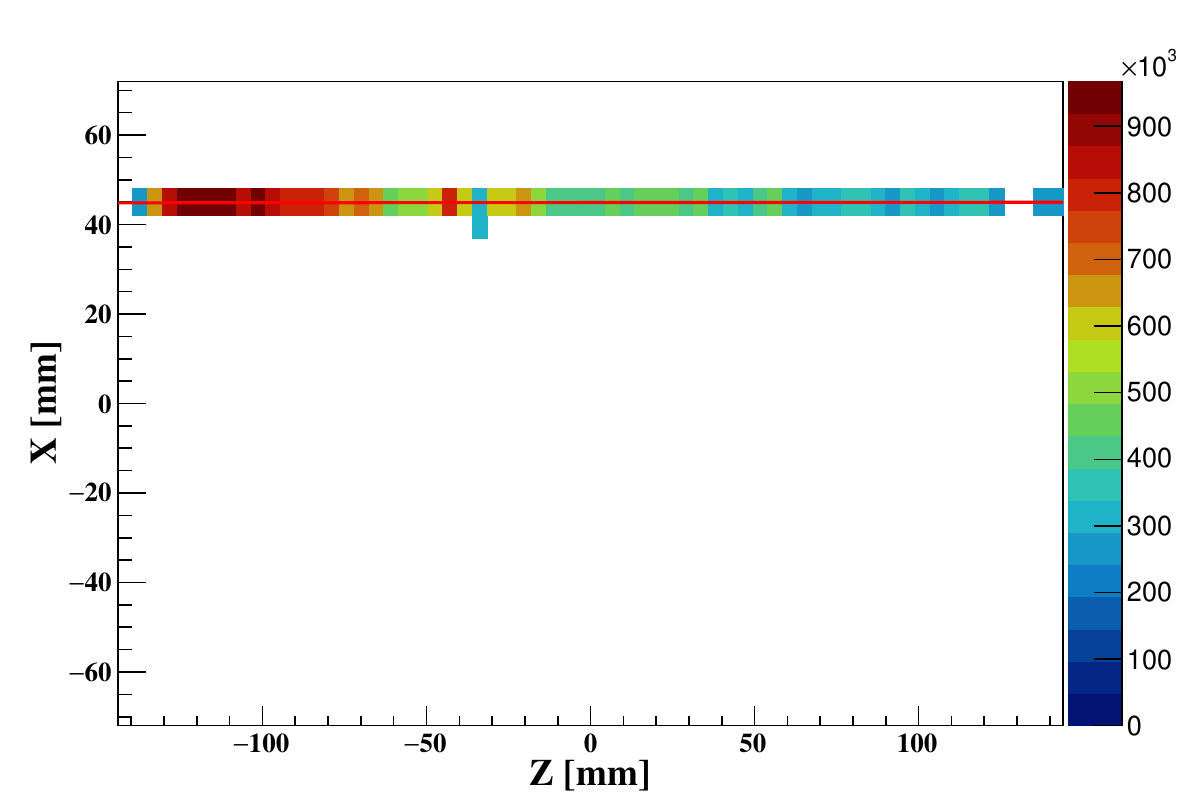}}
     \vspace{3pt}
     \centerline{\includegraphics[width=\textwidth]{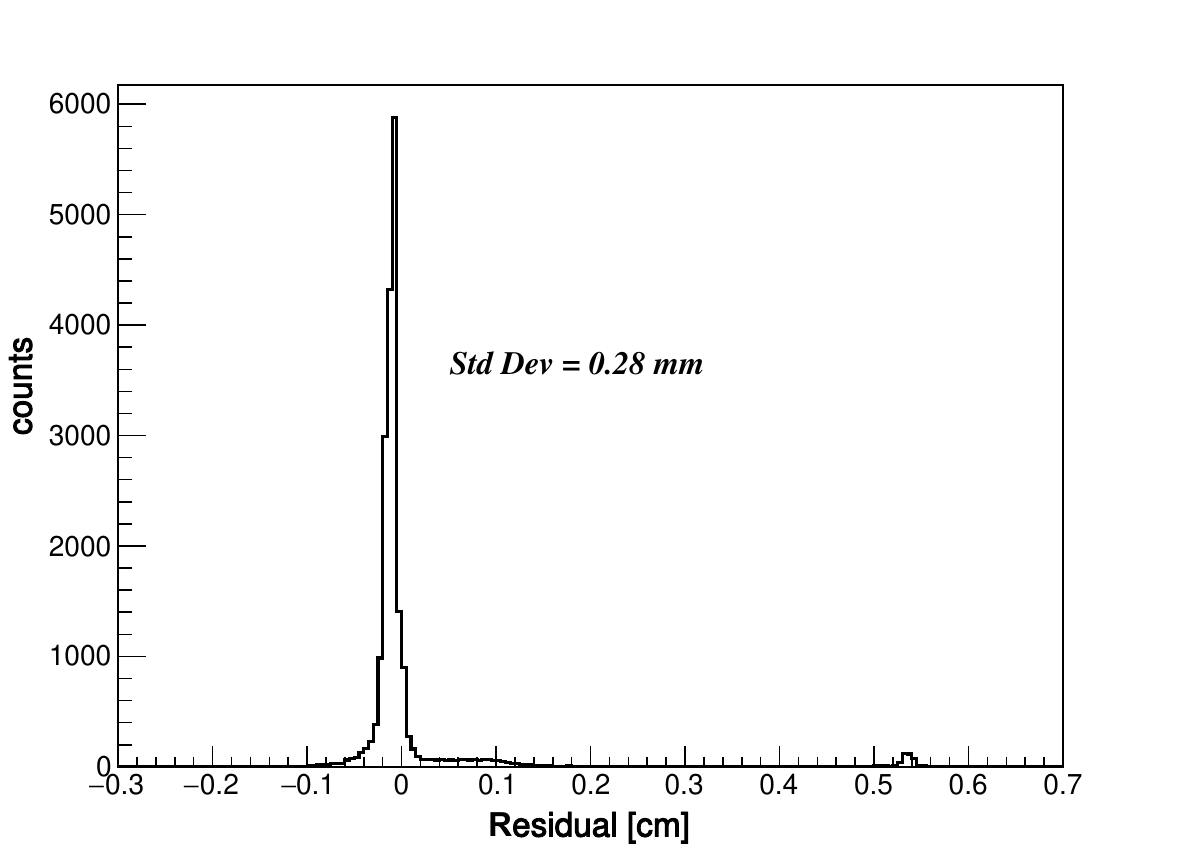}}
  \end{minipage}
  \begin{minipage}{0.48\linewidth}
     \vspace{3pt}
     \centerline{\includegraphics[width=\textwidth]{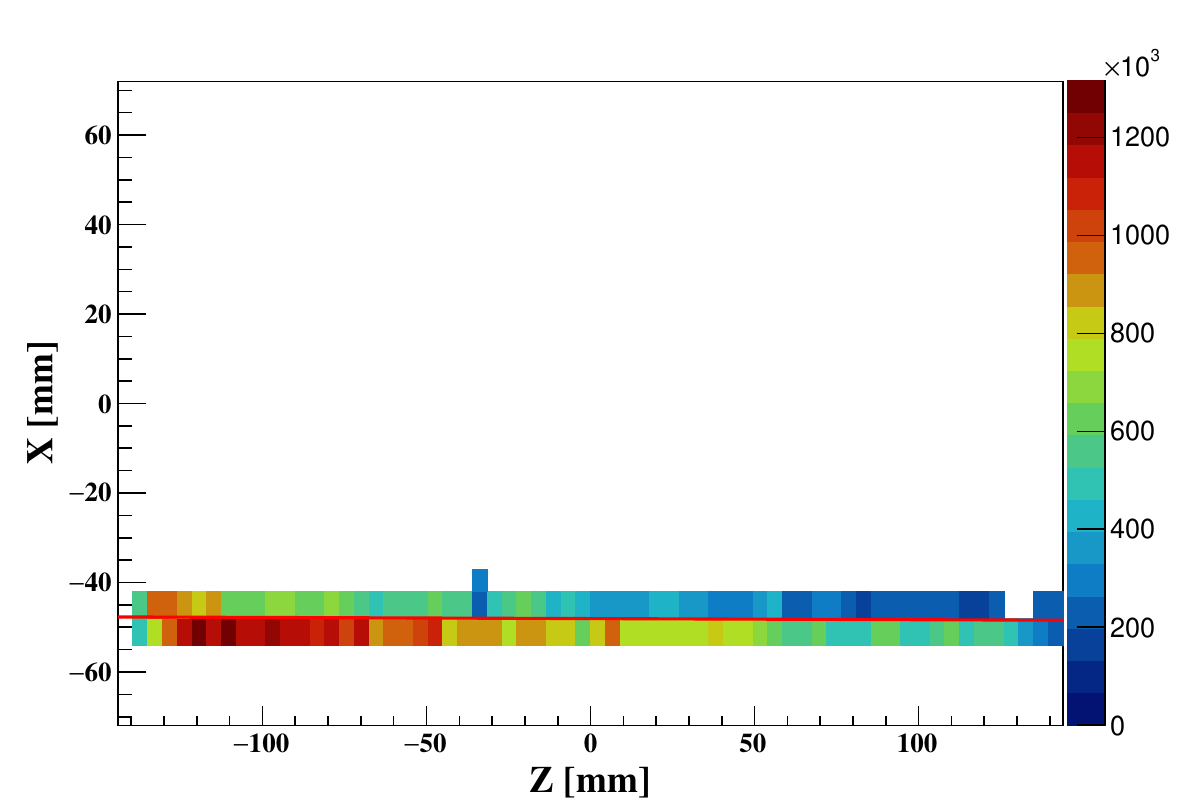}}
     \vspace{3pt}
     \centerline{\includegraphics[width=\textwidth]{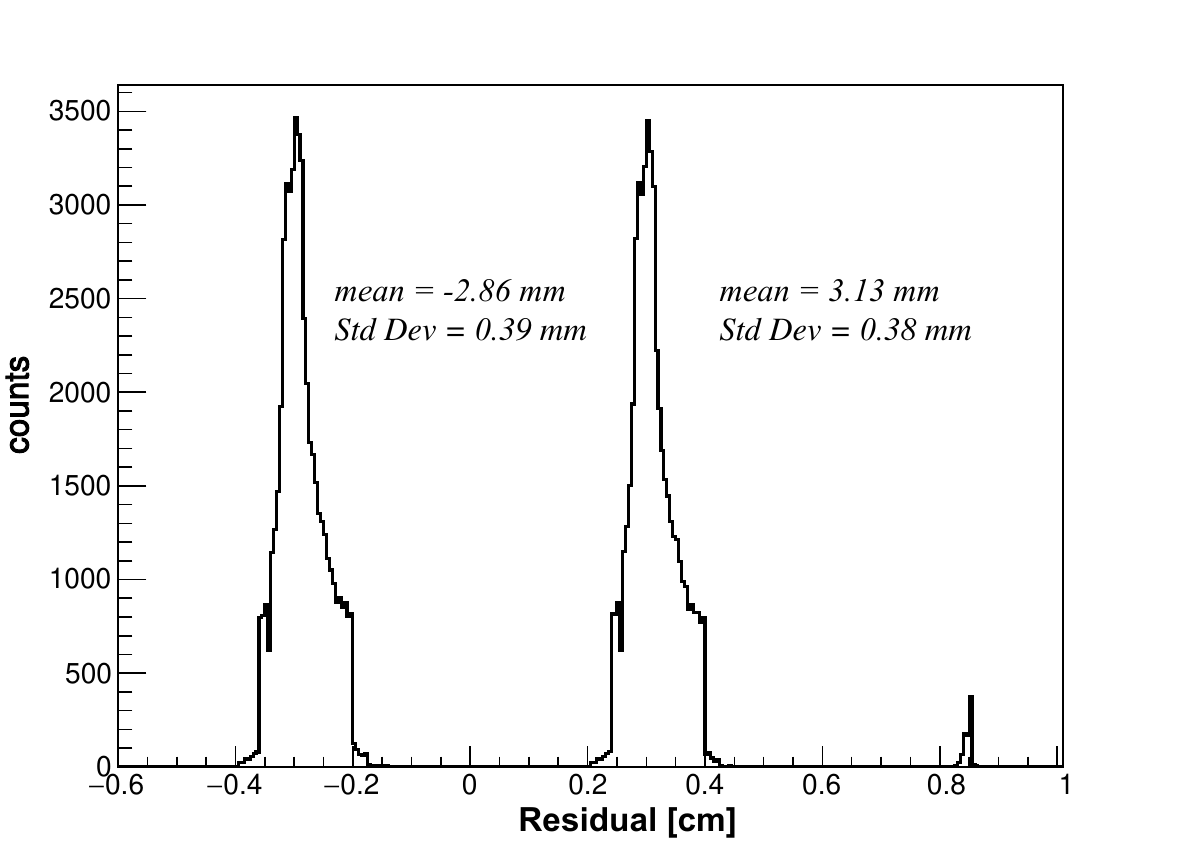}}
 \end{minipage}
 \caption{(Color Online) The top two panels show the recorded two-dimensional trajectories of the laser beam, with the red lines representing two-dimensional weighted linear fits. The lower panel shows the residuals between the fitted curve and the measured positions.}
 \label{fig:field_homogeneity}
\end{figure*}

Ensuring the homogeneity of the electric field is important for achieving accurate 3D trajectory reconstructions. Ideally, the electric field should be perpendicular to the micromesh, have no horizontal components, and maintain a uniform scalar magnitude in the vertical orientation. However, practical considerations introduce several factors that affect the uniformity of the electric field, including wire deformation induced by voltage, feedback from positive ions, and edge effects of the electric field.

To evaluate the homogeneity of the electric field within the sensitive volume, the chamber was filled with 600 mbar of Ar+$\rm iC_4H_{10}$ (93:7) mixed gases. The cathode voltage was set to -1800 V, resulting in a drift electric field strength of 100 V/cm, while the anode avalanche voltage was set to +350 V. 
Considering that distortions are more pronounced at the edges, the uniformity of the electric field at the edge of the sensitive area better reflects the quality of the field cage design and fabrication. Therefore, two trajectories at the edges are provided as an example.
As shown in the figure, the upper two panels show the measured trajectories of the laser beam along the $z$-axis at the edge of the sensitive area, with the red lines representing two-dimensional weighted linear fits. The lower two panels show the results of two measurements, illustrating the distribution of distances between the measured positions and the reconstructed red lines. As shown in the figure, the standard deviation of the residual distributions remains below 0.4 mm for both measurements, indicating that the electric field uniformity within the field cage is satisfactorily maintained.

\subsection{Angular and spatial resolution}\label{sec:angular}

\begin{figure}[!htb]
    \centering
    \includegraphics[width=\linewidth]{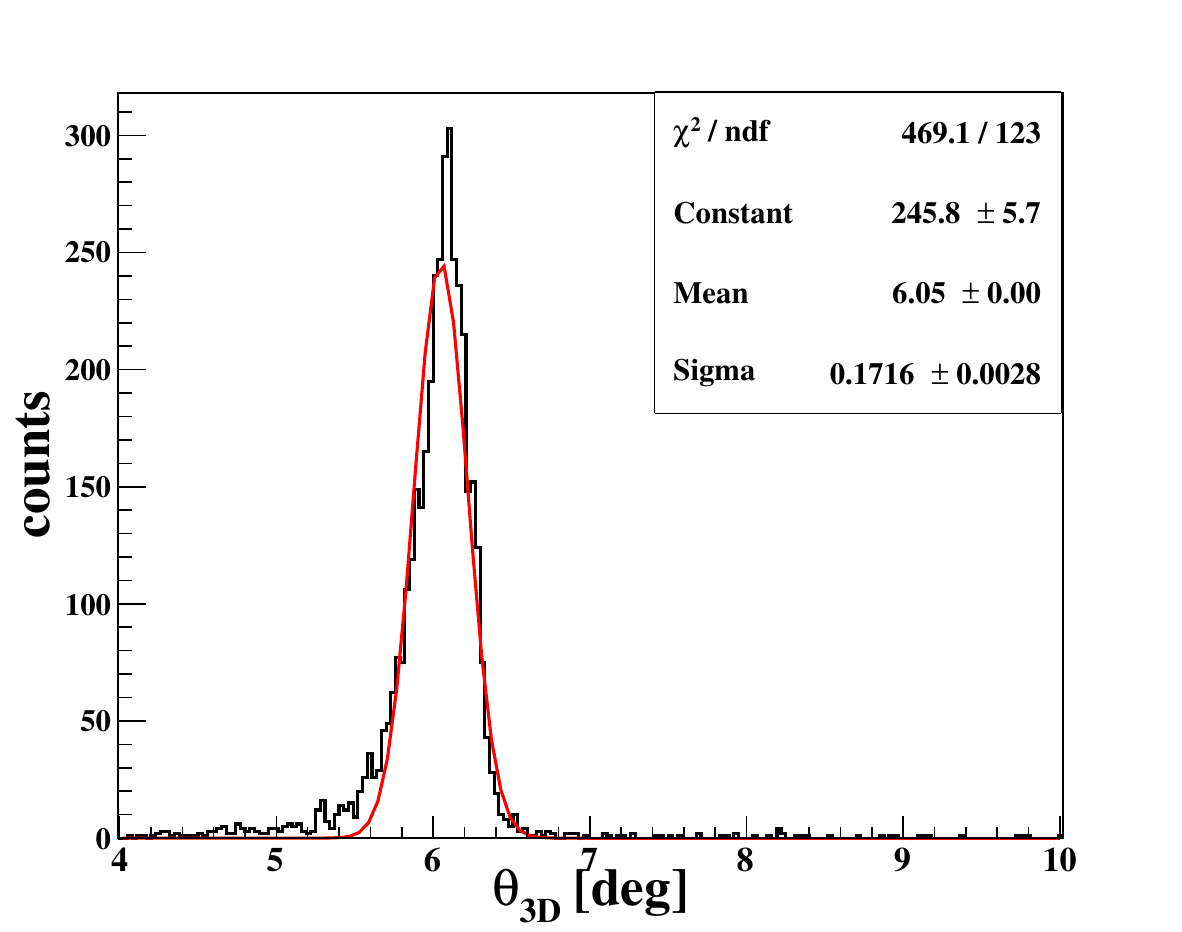}
    \quad
    \includegraphics[width=\linewidth]{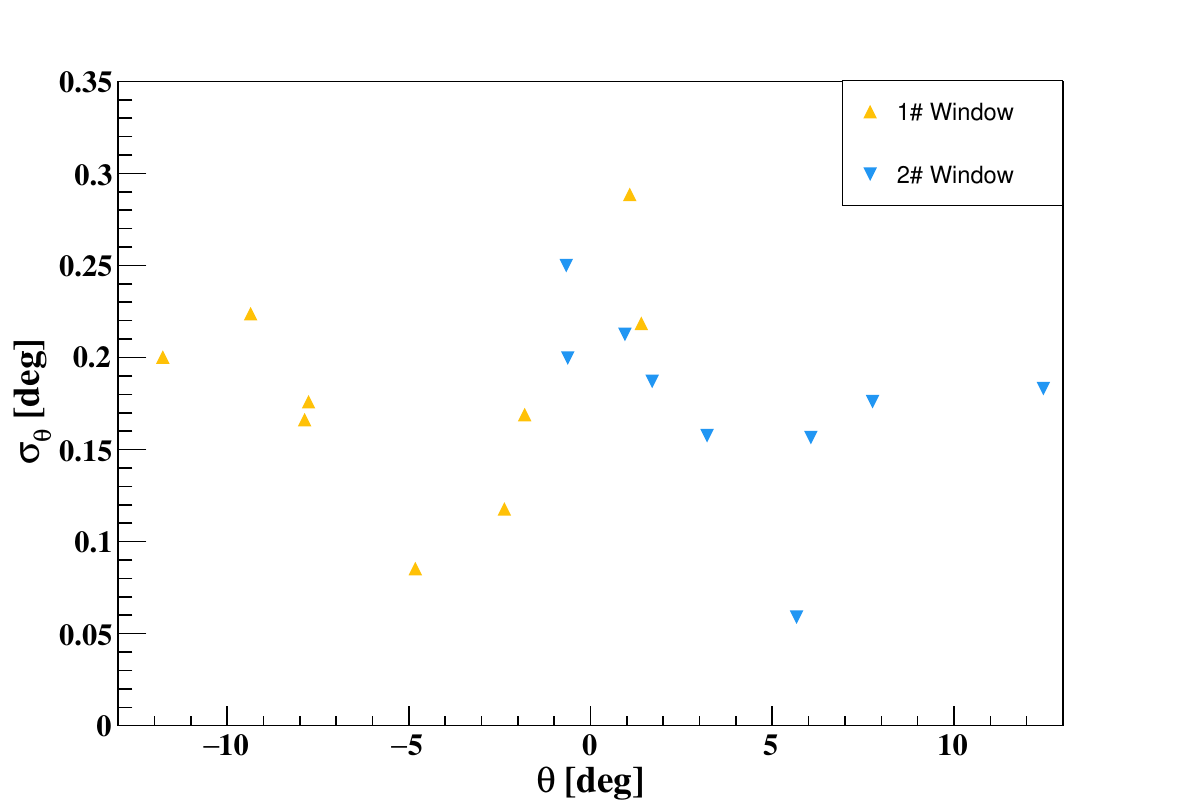}
    \caption{(Color Online) Top panel: The angular distribution of the laser track determined by weighted linear fitting.
    Bottom panel: Intrinsic angular resolutions measured for laser incidences at different angles from two symmetrically positioned quartz windows.}
    \label{fig:laser_3D}
\end{figure}

Since the UV laser light is free of straggling effects, it can be used to test the intrinsic angular and spatial resolution of the detector. In this work, $\theta$ is defined as the angle between the projection of the track on the readout plane and the $z$ axis. The detector was filled with Ar+$\rm iC_4H_{10}$ (93:7) gas mixtures at 400 mbar. During the measurement, the drift field strength was set to 44 V/cm and the avalanche voltage was set to +345 V. \reffig{fig:laser_3D}(top) shows the measured angular resolution of the laser with an incident angle of $6.05^{\circ}$.  And an angular resolution of 0.17 $^{\circ}$ $ (\sigma)$ is achieved in this incident direction. The incident angle of the laser beam is determined by three-dimensional weighted linear fitting. The weighted coefficient is the ratio of the deposited charge to the area of the pixel.

It should be noted that both angular and spatial resolutions depend on the angle of incidence and the pad structure. Therefore, it is necessary to measure the intrinsic angular resolutions at different incident angles, especially for readout pixels of different sizes, as shown in \reffig{fig:micromegas}. In \reffig{fig:laser_3D}(below) the measured intrinsic angular resolution is shown, with laser light incident through two quartz windows at different angles of incidence. A remarkable angular resolution of 0.06$^{\circ}$ ($\sigma_{\theta}$) is achieved at an angle of incidence of 5.6$^{\circ}$, while the worst measured angular resolution remains within 0.30$^{\circ}$ ($\sigma_{\theta}$) when the laser mainly traverses the pixel with a size of 6$\times$4.5 mm$^2$. For the detector with a sensitive length of 288 mm, the corresponding position resolutions (FWHM) can be derived from 288 $\times$2.355$\sigma_{\theta}$. Thus, the optimal position resolution is 0.71 mm and the worst situation is within 3.4 mm.

\subsection{Gain uniformity} \label{sec:gain}

\begin{figure}[!htb]  
   \centering
    \includegraphics[width=\linewidth]{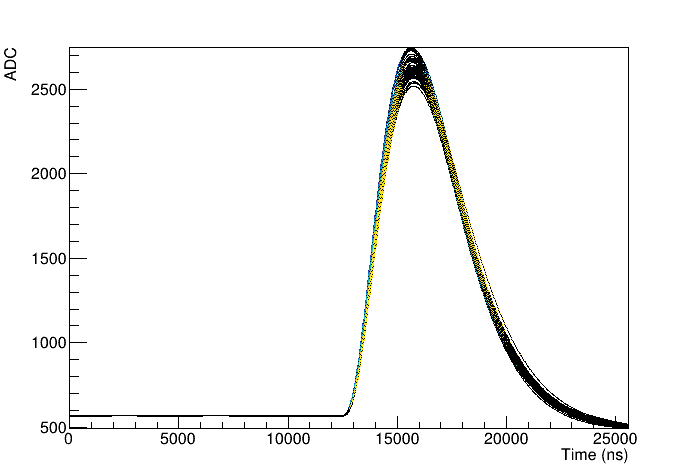}
    \quad
    \includegraphics[width=\linewidth]{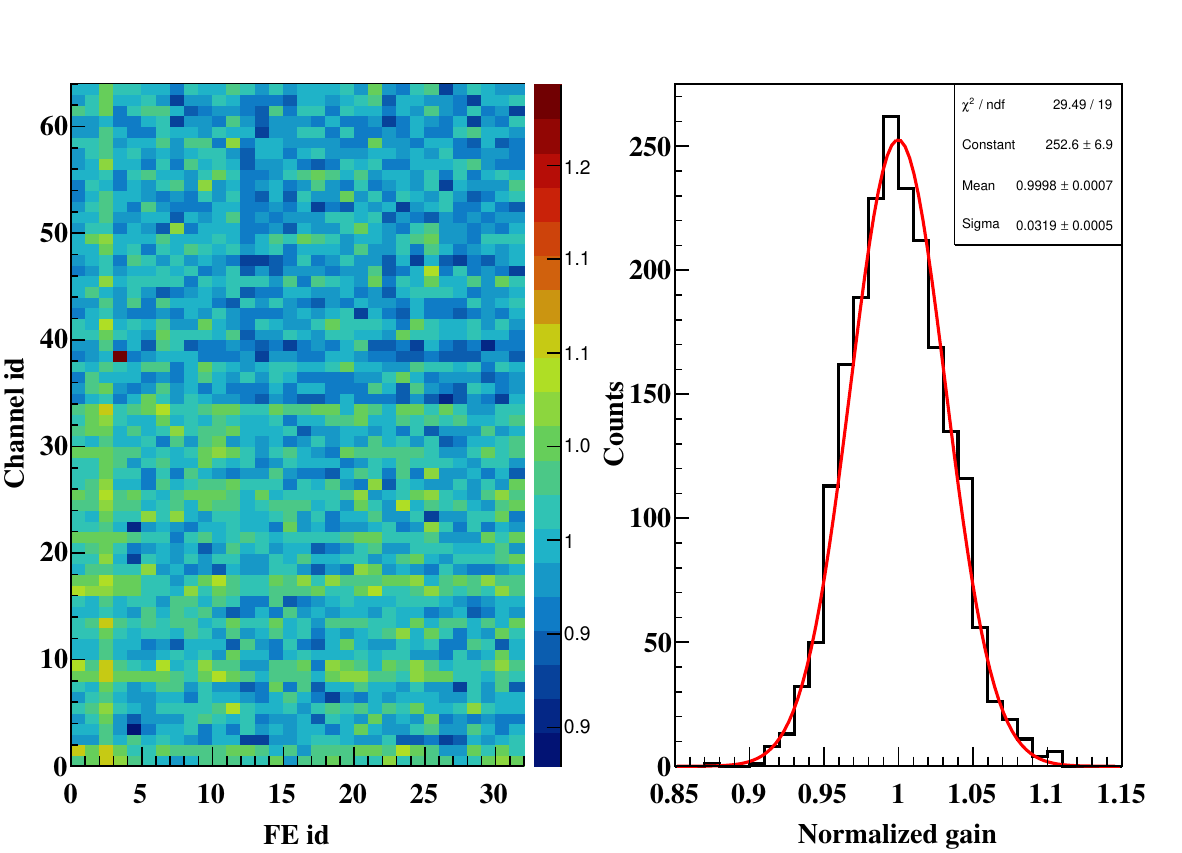}
  \caption{(Color Online) Top panel: 64-channel waveforms acquired from a PAM by injecting a common pulse.
  Bottom panel: (Left) The gain variations of the electronics channel by channel. (Right) The normalized gain distribution of the 2048 channels.}
  \label{fig:electronic_gain}
\end{figure}

\begin{figure}[!htb]
  \centering
  \includegraphics[width=\linewidth]{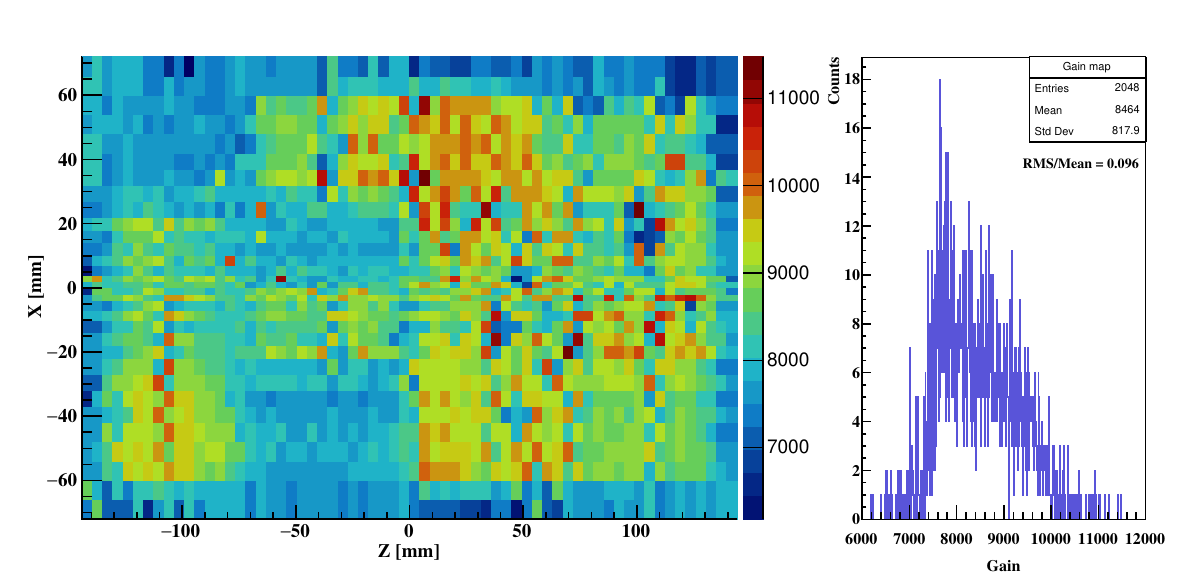}
 \caption{(Color Online) (Left) The gain map of the detector at the avalanche voltage of +465 V. The color scale represents the gain of each pad. (Right) The gain distribution of the 2048 pads.}
 \label{fig:detector_gain}
\end{figure}

The calibration of the gain map plays a critical role in the energy reconstruction of charged particles. In addition, the uniformity of the detector is mainly influenced by two factors. The primary factor is the uniformity of the avalanche gap, while the secondary factor is the uniformity of the gain of the electronics.

First, the gain variations in the electronics are tested by applying a common pulse to the PAM board. In each PAM module, all 64 channels are connected to the Sub-Miniature-A (SMA) connector via 1pF capacitors with 5\% accuracy for gain calibration. The resulting waveform acquired from the 64 channels within a PAM module is visually represented in the top panel of \reffig{fig:electronic_gain}. By adjusting the pulse amplitudes to record the waveform amplitudes for each channel at different collected charge levels, the resulting normalized gain variation distribution for each channel is shown in the bottom panel of \reffig{fig:electronic_gain}. The result shows that the gain variation across the 2048 channels of the electronic system is within 4\% ($\sigma/\mu$).

\begin{figure}[!htb]  
   \centering
    \includegraphics[width=\linewidth]{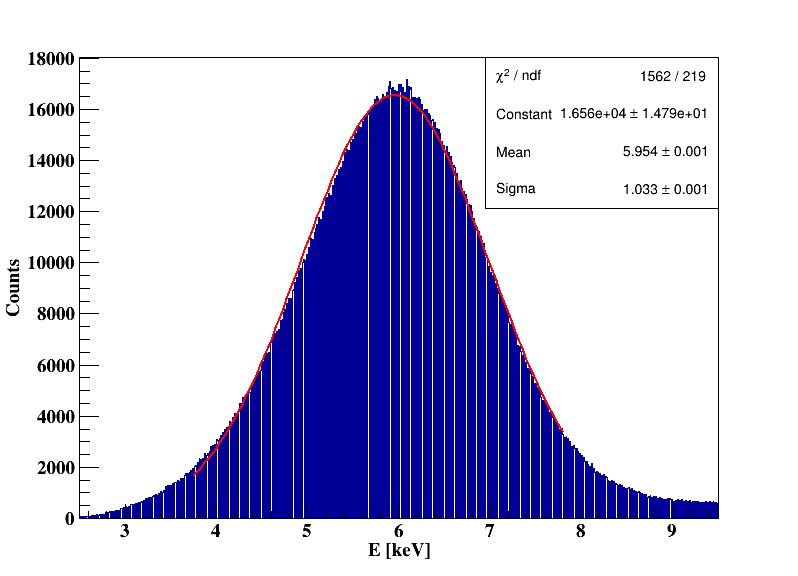}  
    \quad
    \includegraphics[width=\linewidth]{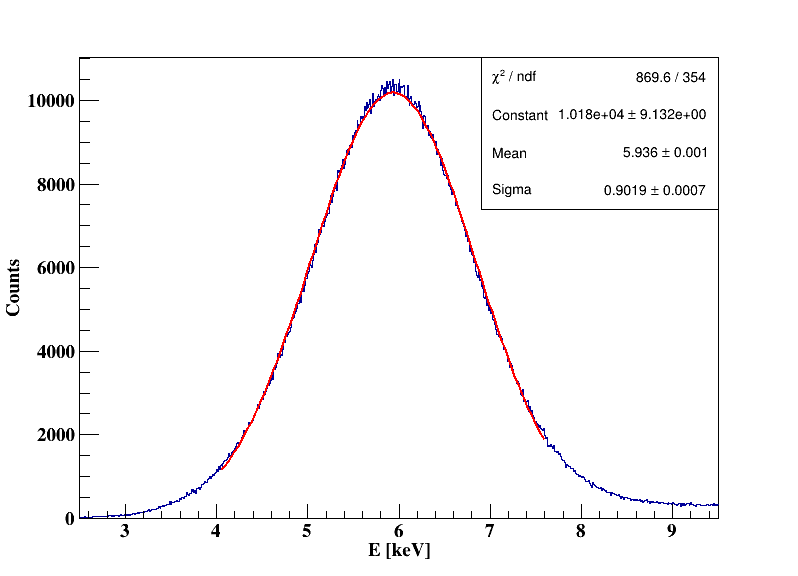}
  \caption{(Color Online) Top Panel: The $^{55}$Fe spectrum without gain map.
  Bottom Panel: The $^{55}$Fe spectrum with the correction of gain map.}
  \label{fig:600mbar_P10_Fe55_Map}
\end{figure}

A $^{55}$Fe x-ray source with an activity of 1.43$\times$10$^5$ becquerels and a diameter of 10 mm was then used to evaluate the gain uniformity of the Micromegas detector. The source was positioned at the center of the pad plane and at a distance of 20 cm from the micromesh. This ensures that every pixel can be illuminated. For the gain uniformity test, the chamber was filled with P10 gas at a pressure of 600 mbar. The anode voltage was set to +465 V, and the field ratio of the avalanche field to the drift field was set to 300, ensuring that more than 95\% of the primary electrons could be collected in the avalanche gap at this field ratio.

\begin{equation}
    \label{eq1}
    \left[\begin{array}{ccccc}
        x_1 & x_2 & \cdots & x_{2048} \\
        x_1^2 & x_2^2 & \cdots & x_{2048}^2 \\
        \cdots & \cdots & \vdots & \cdots \\
        x_1^m & x_2^m & \cdots & x_{2048}^m
    \end{array}\right]\left[\begin{array}{c}
    \theta_1 \\
    \theta_2 \\
    \vdots \\
    \theta_{2048}
    \end{array}\right]=\left[\begin{array}{c}
    y_1 \\
    y_2 \\
    \vdots \\
    y_m
    \end{array}\right]
\end{equation}

Due to the transverse diffusion of electrons, a typical 5.9 keV X-ray event consists of 2-7 pixels. To calculate the gain value for each pixel, the multiple linear regression method was used to analyze the data. In \refequ{eq1}, the input matrix on the left represents the deposited charge of each pixel for each $^{55}$Fe event. The middle parameter vector $\mathbf{\theta}$ represents the gain map of the detector, and the output vector $\mathbf{Y}$ represents the primary electrons ionized by the 5.9 keV X-ray. According to the method of least squares, the parameter vector can be calculated from $\mathbf{\theta}=(\mathbf{X}^\mathrm{T} \mathbf{X})^{-1}\cdot \mathbf{X}^\mathrm{T}\mathbf{Y} $. The resulting gain map, shown in \reffig{fig:detector_gain} (left), revealed a gain variation of about 9.6\% (RMS/mean) over 2048 channels. This result implies that the fluctuations in detector gain are primarily caused by the non-uniformity of the detector's avalanche gaps.

To validate the results obtained, the gain map was adjusted to correct the $^{55}$Fe spectrum measured at the anode. 
The comparison of the spectrum before and after correction is shown in \reffig{fig:600mbar_P10_Fe55_Map}.
The spectrum before correction was derived assuming uniform pixel gains. 
It can be seen that after applying the gain map, the width ($\sigma$) of the $^{55}$Fe spectrum decreased from 1.03 keV to 0.90 keV. This result indicates that the gain map obtained by multiple linear regression has some reliability.

\subsection{Energy resolution} \label{sec:ER}

\begin{figure}[!htb]
  \centering
  \includegraphics[width=\linewidth]{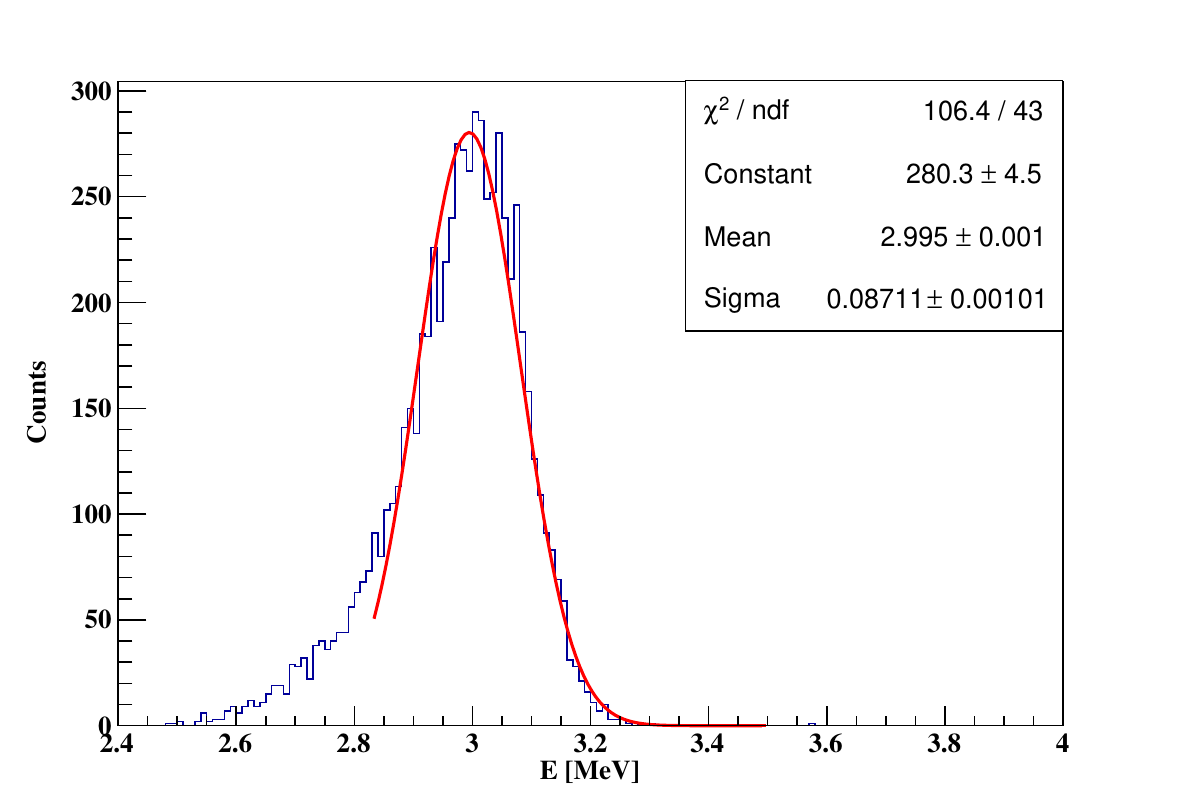}
 \caption{(Color Online) The energy spectrum obtained from $^{241}$Am source.}
 \label{fig:241Am_E}
\end{figure}

The energy resolution of the detector was further investigated using a $^{241}$Am alpha source located 39 mm from the sensitive region. The detector was filled with 600 mbar of P10 gas, and the voltages applied to the anode and cathode were set to +405 V and -2430 V, respectively. The total charge deposited by the alpha particles was obtained on an event-by-event basis by summing all the individual charges collected on each pad. To minimize the effect of the dead zone between the alpha source and the sensitive area on the energy resolution, it is necessary to collimate the alpha source by restricting the emission angle of alpha to within $\pm 5^{\circ}$ during data analysis. The resulting energy spectrum for the alpha source is shown in \reffig{fig:241Am_E}. Fitting the spectrum with single Gaussian distributions yielded an energy resolution of 6.85\% for alpha particles with a deposited energy of 3.0 MeV.

\section{Summary} \label{Summary}
A new 2048-channel prototype active target detector system, called fMeta-TPC, has been designed and constructed for low-energy nuclear experiments. 
In this work, the resistive micromegas with an avalanche gap of 100 $\mu$m is used for signal readout. As verified by the $^{55}$Fe X-ray source, the gain uniformity of the detector is about 10\% (RMS/mean), and the contribution of electronic gain fluctuations is within 4\% ($\sigma/\mu$). The energy resolution obtained from the total charge collected on the pad plane was deduced to be 6.85\% for 3.0 MeV alpha particles.

Considering the characteristics of photodisintegration reactions, the readout plate is divided into 2048 rectangular pixels of unequal size. In terms of angular resolution, it was tested with laser light at different injection angles to evaluate the effect of pad size on angular resolution. The results show that the readout board can achieve a remarkable angular resolution of 0.06$^{\circ}$ ($\sigma_{\theta}$), and the worst angular resolution measured is within 0.30$^{\circ}$ ($\sigma_{\theta}$). In addition, the electron drift velocity and the homogeneity of the drift field by laser light were also tested in this work. The results show that the homogeneity of the drift field is satisfactorily maintained in the sensitive volume. And the measured electron drift velocity is in good agreement with other experimental results and theoretical calculations.

For future low-pressure experiments, it is planned to upgrade the detection system by using micromegas with a double micromesh gaseous structure (DMM) \cite{bib:42,bib:43} or micromegas with a larger avalanche gap. In addition, the first commissioning run of the $^7\mathrm{Li}(\gamma,t)^4\mathrm{He}$ ground-state cross section has been performed at SLEGS and the data are currently being processed.

\section*{CrediT authorship contribution statement}
Huangkai Wu: Methodology, Software, Formal analysis, Investigation, Data curation, Writing – original draft, Writing – review \& editing. 
Xiyang Wang: Formal analysis, Validation, Investigation, Data curation. 
Yumiao Wang: Investigation, Validation, Formal analysis. 
Youjing Wang: Investigation, Validation, Formal analysis.
Deqing Fang: Investigation, Resources. 
Wanbing He: Software, Investigation. 
Weihu Ma: Investigation. 
Xiguang Cao: Investigation. 
Changbo Fu: Conceptualization, Methodology, Validation, Formal analysis, Investigation, Writing – origin draft, Writing – review \& editing, Supervision, Project administration. 
Xiangai Deng: Methodology, Validation, Investigation, Resources, Writing – review \& editing, Supervision, Project administration. 
Yugang Ma: Investigation, Resources, Writing – review \& editing, Supervision. 


\end{document}